\documentclass[aps,prl,reprint,twocolumn,nopacs,superscriptaddress]{revtex4}
\usepackage{amsmath}
\usepackage{amssymb}
\usepackage{graphicx}
\usepackage{hyperref}
\newcommand{\beq}{\begin{equation*}}
\newcommand{\eeq}{\end{equation*}}

\begin{document}

\title{Criteria for directly detecting topological Fermi arcs in Weyl semimetals}

\author{Ilya Belopolski\footnote{These authors contributed equally to this work.}} \email{ilyab@princeton.edu} \affiliation{Laboratory for Topological Quantum Matter and Spectroscopy (B7), Department of Physics, Princeton University, Princeton, New Jersey 08544, USA}
\author{Su-Yang Xu$^*$} \affiliation{Laboratory for Topological Quantum Matter and Spectroscopy (B7), Department of Physics, Princeton University, Princeton, New Jersey 08544, USA}
\author{Daniel S. Sanchez$^*$} \affiliation{Laboratory for Topological Quantum Matter and Spectroscopy (B7), Department of Physics, Princeton University, Princeton, New Jersey 08544, USA}
\author{Guoqing Chang} \affiliation{Centre for Advanced 2D Materials and Graphene Research Centre, National University of Singapore, 6 Science Drive 2, 117546, Singapore} \affiliation{Department of Physics, National University of Singapore, 2 Science Drive 3, 117542, Singapore}
\author{Cheng Guo} \affiliation{International Center for Quantum Materials, Peking University, Beijing 100871, China}
\author{Madhab Neupane} \affiliation{Condensed Matter and Magnet Science Group, Los Alamos National Laboratory, Los Alamos, NM 87545, USA} \affiliation{Department of Physics, University of Central Florida, Orlando, FL 32816, USA}
\author{Hao Zheng} \affiliation{Laboratory for Topological Quantum Matter and Spectroscopy (B7), Department of Physics, Princeton University, Princeton, New Jersey 08544, USA}
\author{Chi-Cheng Lee} \affiliation{Centre for Advanced 2D Materials and Graphene Research Centre, National University of Singapore, 6 Science Drive 2, 117546, Singapore} \affiliation{Department of Physics, National University of Singapore, 2 Science Drive 3, 117542, Singapore}
\author{Shin-Ming Huang} \affiliation{Centre for Advanced 2D Materials and Graphene Research Centre, National University of Singapore, 6 Science Drive 2, 117546, Singapore} \affiliation{Department of Physics, National University of Singapore, 2 Science Drive 3, 117542, Singapore}
\author{Guang Bian} \affiliation{Laboratory for Topological Quantum Matter and Spectroscopy (B7), Department of Physics, Princeton University, Princeton, New Jersey 08544, USA}
\author{Nasser Alidoust} \affiliation{Laboratory for Topological Quantum Matter and Spectroscopy (B7), Department of Physics, Princeton University, Princeton, New Jersey 08544, USA}
\author{Tay-Rong Chang} \affiliation{Laboratory for Topological Quantum Matter and Spectroscopy (B7), Department of Physics, Princeton University, Princeton, New Jersey 08544, USA} \affiliation{Department of Physics, National Tsing Hua University, Hsinchu 30013, Taiwan}
\author{BaoKai Wang} \affiliation{Centre for Advanced 2D Materials and Graphene Research Centre, National University of Singapore, 6 Science Drive 2, 117546, Singapore} \affiliation{Department of Physics, National University of Singapore, 2 Science Drive 3, 117542, Singapore} \affiliation{Department of Physics, Northeastern University, Boston, Massachusetts 02115, USA}
\author{Xiao Zhang} \affiliation{International Center for Quantum Materials, Peking University, Beijing 100871, China}
\author{Arun Bansil} \affiliation{Department of Physics, Northeastern University, Boston, Massachusetts 02115, USA}
\author{Horng-Tay Jeng} \affiliation{Department of Physics, National Tsing Hua University, Hsinchu 30013, Taiwan} \affiliation{Institute of Physics, Academia Sinica, Taipei 11529, Taiwan}
\author{Hsin Lin} \affiliation{Centre for Advanced 2D Materials and Graphene Research Centre, National University of Singapore, 6 Science Drive 2, 117546, Singapore} \affiliation{Department of Physics, National University of Singapore, 2 Science Drive 3, 117542, Singapore}
\author{Shuang Jia} \affiliation{International Center for Quantum Materials, Peking University, Beijing 100871, China}
\author{M. Zahid Hasan} \email{mzhasan@princeton.edu} \affiliation{Laboratory for Topological Quantum Matter and Spectroscopy (B7), Department of Physics, Princeton University, Princeton, New Jersey 08544, USA}


\begin{abstract}
The recent discovery of the first Weyl semimetal in TaAs provides the first observation of a Weyl fermion in nature and demonstrates a novel type of anomalous surface state, the Fermi arc. Like topological insulators, the bulk topological invariants of a Weyl semimetal are uniquely fixed by the surface states of a bulk sample. Here, we present a set of distinct conditions, accessible by angle-resolved photoemission spectroscopy (ARPES), each of which demonstrates topological Fermi arcs in a surface state band structure, with minimal reliance on calculation. We apply these results to TaAs and NbP. For the first time, we rigorously demonstrate a non-zero Chern number in TaAs by counting chiral edge modes on a closed loop. We further show that it is unreasonable to directly observe Fermi arcs in NbP by ARPES within available experimental resolution and spectral linewidth. Our results are general and apply to any new material to demonstrate a Weyl semimetal.
\end{abstract}

\date{\today}
\maketitle

\begin{figure}
\centering
\includegraphics[width=8.5cm, trim={30 45 80 80}, clip]{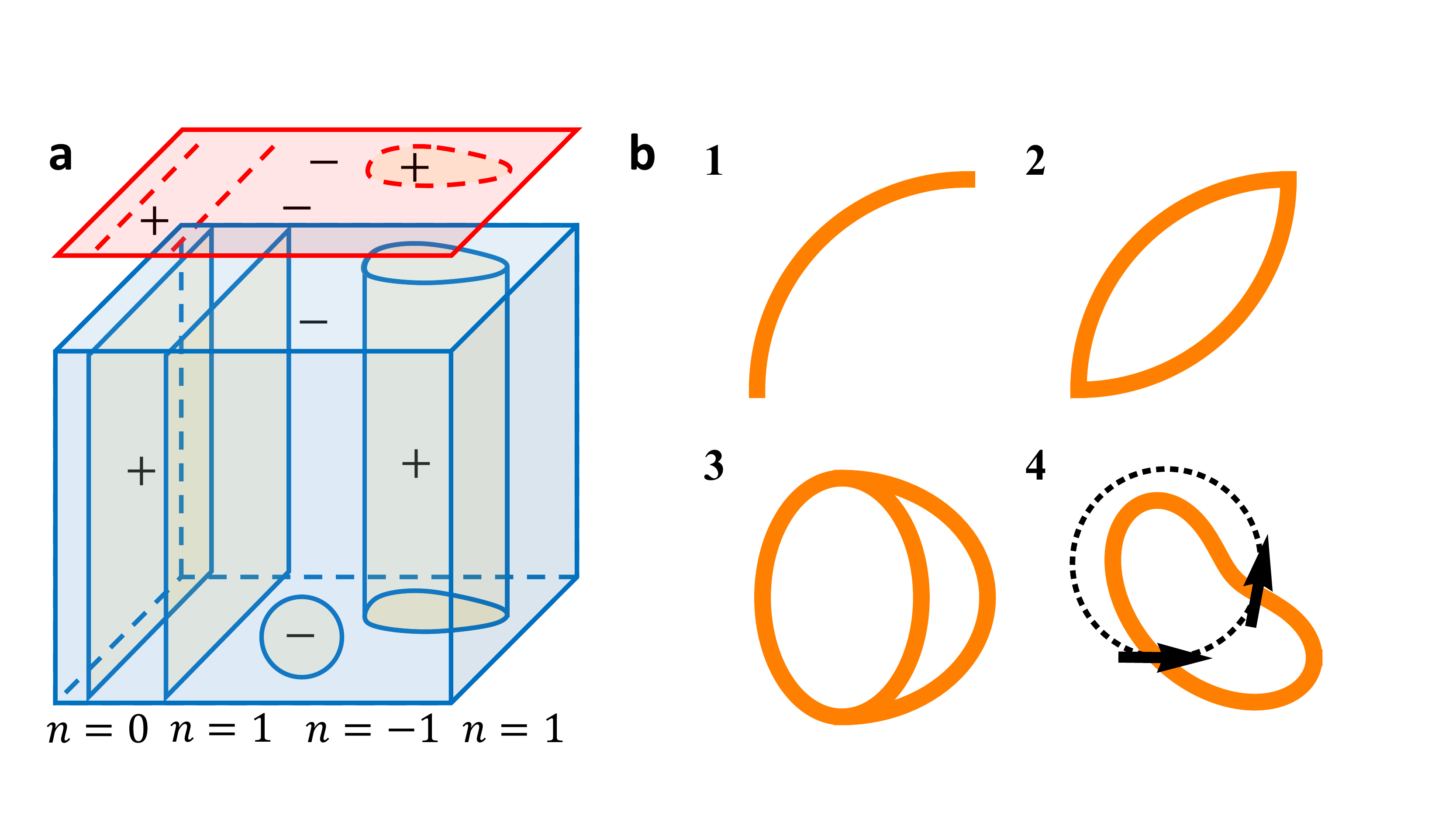}
\caption{\label{Fig1}\textbf{Four criteria for Fermi arcs.} (a) A bulk (blue) and surface (red) Brillouin zone with four Weyl points ($\pm$) and various 2D manifolds with Chern number $n$. Sweeping a plane through a Weyl point changes $n$. Also, any closed loop on the surface hosts chiral edge modes protected by the enclosed bulk chiral charge. For instance, a closed loop enclosing a $+$ Weyl point will host one right-moving chiral edge mode \cite{Hosur, Pyrochlore, Vish}. (b) The criteria: (1) A disjoint contour. (2) A closed contour with a kink. (3) No kinks within experimental resolution, but an odd set of closed contours. (4) An even number of contours without kinks, but net non-zero chiral edge modes.}
\end{figure}

A Weyl semimetal is a crystal which hosts Weyl fermions as emergent quasiparticles \cite{Weyl, Herring, Abrikosov, Nielsen, Volovik, Murakami, Pyrochlore, Multilayer, Hosur, Vish}. Although Weyl fermions are well-studied in quantum field theory, they have not been observed as a fundamental particle in nature. The recent experimental observation of Weyl fermions as electron quasiparticles in TaAs offers a beautiful example of emergence in science \cite{TaAsThyUs, TaAsThyThem, TaAsUs, LingLu, TaAsThem, TaAsChen, TaAsNodesDing, NbAs, TaPUs, TaPThem, HaoNbP}. Weyl semimetals also give rise to a topological classification closely related to the Chern number of the integer quantum Hall effect \cite{Pyrochlore, Vish, Hosur, Bernevig}. In the bulk band structure of a three-dimensional sample, Weyl fermions correspond to points of accidental degeneracy, Weyl points, between two bands \cite{Murakami, Multilayer}. The Chern number on a two-dimensional slice of the Brillouin zone passing in between Weyl points can be non-zero, as illustrated in Fig. \ref{Fig1}(a) \cite{Pyrochlore, Vish, Hosur}. Exactly as in the quantum Hall effect, the Chern number in a Weyl semimetal protects topological boundary modes. However, the Chern number changes when the slice is swept through a Weyl point, so the chiral edge modes associated with each slice terminate in momentum space at the locations of Weyl points, giving rise to Fermi arc surface states. This bulk-boundary correspondence makes it possible to demonstrate that a material is a Weyl semimetal by measuring Fermi arc surface states alone. As a result, to show novel Weyl semimetals, it is crucial to understand the possible signatures of Fermi arcs in a surface state band structure.

\begin{figure*}
\centering
\includegraphics[width=15cm, trim={20 20 10 30}, clip]{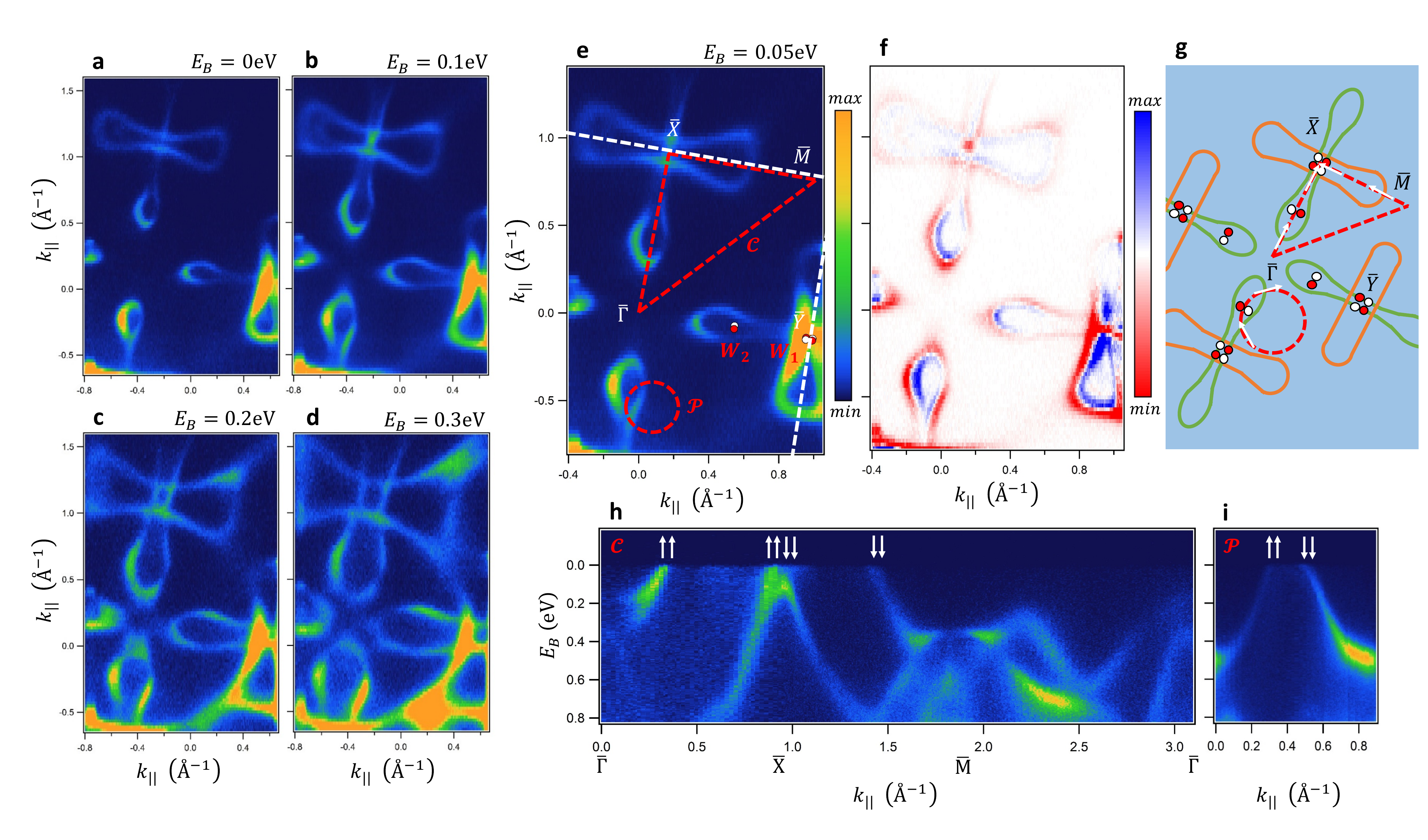}
\caption{\label{Fig2}\textbf{Surface states of NbP by ARPES.} (a)-(d) Constant energy cut by vacuum ultraviolet APRES, at incident photon energy $h\nu = 30$eV, on the (001) surface of NbP. (e) Same as (a)-(d), at $E_B = 0.05$ eV, with Weyl points marked and the two paths $\mathcal{C}$ and $\mathcal{P}$ on which we measure Chern numbers. (f) The difference of ARPES spectra at $E_B = 0.05$ eV and $E_B = 0.1$ eV, illustrating the direction of the Fermi velocity all around the lollipop and peanut pockets. The blue contour is always inside the red contour, indicating that the sign of the Fermi velocity is the same going around each contour. This result excludes the possibility that the lollipop actually consists of Fermi arcs attached to the $W_2$. (g) Cartoon summarizing the band structure. (h) Band structure along $\mathcal{C}$, with chiralities of edge modes marked by the arrows. We associate one arrow to each spinful crossing, even where we cannot observe spin splitting due to the weak SOC of NbP. There are the same number of arrows going up as down, so the Chern number is zero. (i) Same as (h) but along $\mathcal{P}$. Again, the Chern number is zero.}
\end{figure*}

\begin{figure*}
\centering
\includegraphics[width=15cm, trim={120 150 60 90}, clip]{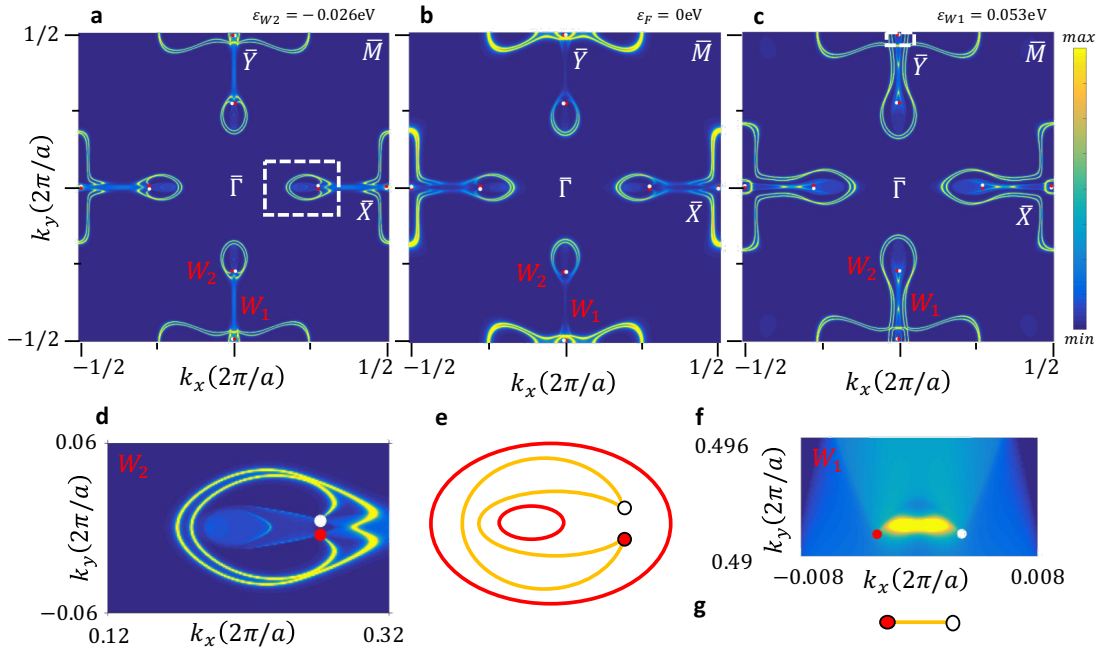}
\caption{\label{Fig3}\textbf{Numerical calculation of Fermi arcs in NbP.} First-principles band structure calculation of the (001) surface states of NbP at (a) the energy of the $W_2$, above the Fermi level, (b) the Fermi level and (c) the energy of the $W_1$, below the Fermi level. In (a) we see (1) a small set of surface states near the $W_2$ and (2) larger surface states near $\bar{X}$ and $\bar{Y}$. Near the energy of (b) there is a Lifshitz transition between the surface states at (1) and (2), giving rise to lollipop and peanut-shaped pockets. At (c) we see that the lollipop and peanut pockets enlarge, so they are hole-like. We also observe short Fermi arcs connecting the $W_1$. The numerical calculation shows excellent overall agreement with our ARPES spectra, suggesting that NbP is a Weyl semimetal. (d) Zoom-in of the surface states around the $W_2$, indicated by the white box in (a). We find two Fermi arcs and two trivial closed contours, illustrated in (e). (f) Zoom-in of the surface states around $W_1$, indicated by the white box at the top of (c). We find one Fermi arc, illustrated in (g).}
\end{figure*}

\begin{figure*}
\centering
\includegraphics[width=17cm, trim={30 180 20 155}, clip]{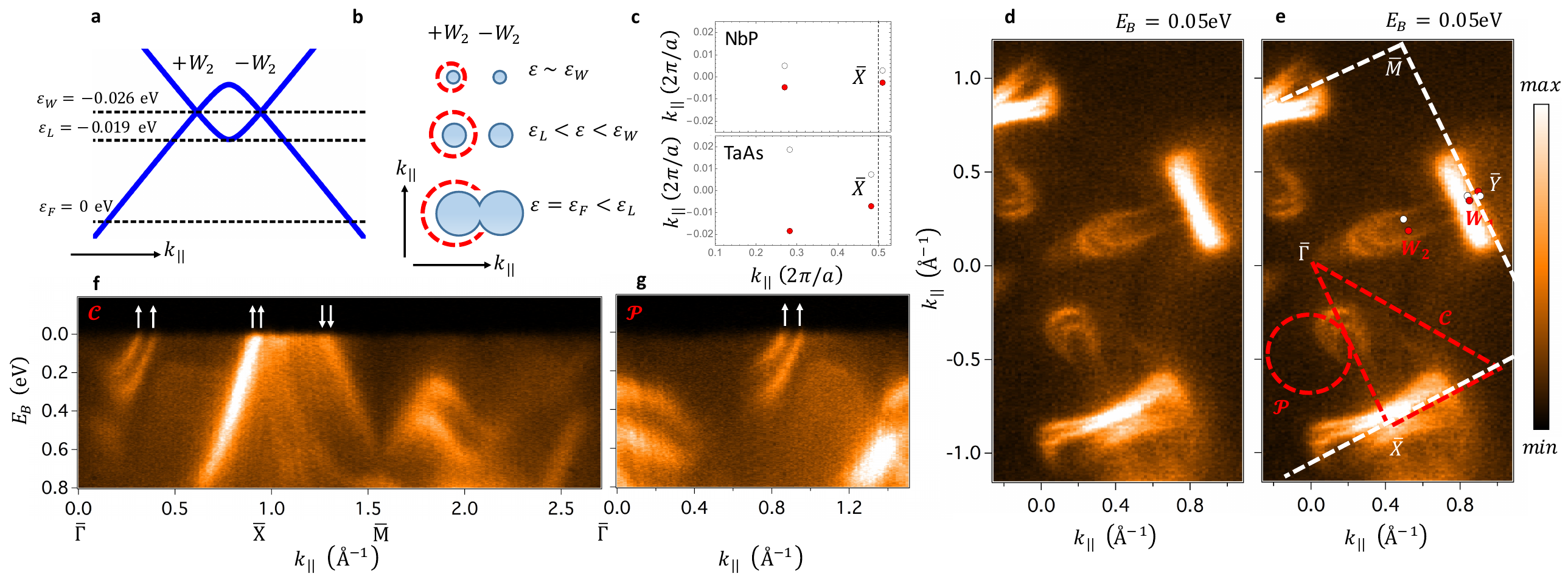}
\caption{\label{Fig4}\textbf{Comparison of NbP and TaAs.} (a) Energies of the $W_2$, their Lifshitz transition and the Fermi level from the numerical calculation. One consequence of the small spin-splitting is that the Lifshitz transition between the $W_2$ is only $\sim 0.007$ eV below the energy of the $W_2$ and $\sim 0.019$ eV above the Fermi level. (b) Because $\varepsilon_F < \varepsilon_L$, it makes no sense to calculate the Chern number on $\mathcal{C}$ and $\mathcal{P}$, because the system on that cut is not an insulator, see the interrupted dotted red line in the last row of (b). (c) Plot of the positions of the Weyl point projections in NbP and TaAs arising from one Dirac line (see $\S 1$, SI). We see that the separation of the Weyl points is $\sim 4$ times larger in TaAs due to the large SOC. (d) Fermi surface by vacuum ultraviolet APRES, at incident photon energy $h\nu = 90$eV, on the (001) surface of TaAs. (e) Same as (d), but with Weyl points marked and two paths $\mathcal{C}$ and $\mathcal{P}$. (f) Band structure along $\mathcal{C}$, with chiralities of edge modes marked by the arrows. The net Chern number appears to be $+2$, inconsistent with the calculation. This can be explained by considering the small separation between the $W_1$. (g) Same as (f) but along $\mathcal{P}$. The path encloses only the well-spaced Weyl points and we find a Chern number $+2$, consistent with the calculation.}
\end{figure*}

Here, we present four distinct signatures of Fermi arcs, all in principle experimentally accessible in an ARPES measurement of a surface state band structure. Each signature alone, observed in any set of surface state bands, is sufficient to show that a material is a Weyl semimetal. Then, we apply these criteria to TaAs and NbP. We find that we can rigorously demonstrate a Weyl semimetal in TaAs by directly measuring a non-zero Chern number in ARPES. For NbP, we find that it is unreasonable to directly observe Fermi arcs within experimental resolution because the spin-orbit coupling is too weak.

Any surface state band structure which satisfies any one of the following four conditions contains topological Fermi arcs and is necessarily a Weyl semimetal:

\begin{enumerate}
\item \label{dis} \textit{Disjoint arc}: Any surface state constant-energy contour with an open curve is a Fermi arc and demonstrates a Weyl semimetal.
\item \label{kink} \textit{Kink off a rotation axis}: A Weyl point is characterized by chiral charge $n$, equal to the Chern number on a small spherical manifold enclosing the Weyl point in the bulk Brillouin zone, illustrated by the small sphere in Fig. \ref{Fig1}(a) \cite{Vish, Hosur, Pyrochlore}. For a Weyl point of chiral charge $|n| > 1$ or if multiple Weyl points project onto the same point of the surface Brillouin zone, there may not be a disjoint constant-energy contour because multiple arcs will emanate from the same Weyl point projection. However, the arcs will generically attach to the Weyl point at different slopes, giving rise to a kink in the constant-energy contour. Moreover, such a kink can only arise from the attachment of two Fermi arcs. A kink on the projection of a rotation axis may arise in a topological Dirac semimetal \cite{Na3Bi}. Off a rotation axis, such a kink guarantees a Weyl semimetal.
\item \label{odd} \textit{Odd number of curves}: For projected chiral charge $|n| > 1$, the constant-energy contours may consist entirely of closed curves and the kink may be below experimental resolution, so the constant-energy contour appears everywhere smooth. However, if $|n|$ is odd, the constant-energy contour will consist of an odd number of curves, so at least one curve must be a Fermi arc.
\item \label{Chern} \textit{Non-zero Chern number}: Consider any closed loop in the surface Brillouin zone where the bulk band structure is everywhere gapped and, at some energy, add up the signs of the Fermi velocities of all surface states along this loop, with $+1$ for right movers and $-1$ for left movers. The sum is the projected chiral charge enclosed in the curve, corresponding to a Chern number on a bulk \cite{Pyrochlore, Vish, Hosur}. A non-zero sum on at least one loop shows a Weyl semimetal, provided the loop is chosen to be contractible on the torus of the surface Brillouin zone.
\end{enumerate}

\noindent Note that while (\ref{dis}), (\ref{kink}) and (\ref{odd}) describe properties of a constant-energy slice of the Fermi surface, the counting argument (\ref{Chern}) requires a measurement of the dispersion. We note also that criterion (\ref{Chern}) allows us to determine all bulk topological invariants and Weyl points of a material by studying only its surface states.

In the rest of this Letter, we apply these criteria to TaAs and NbP. We find that criterion (\ref{Chern}) shows a Weyl semimetal in TaAs, but that all criteria fail for NbP because the spin-orbit coupling is too weak and the Fermi level is too low. We also present a calculation of NbP and show that it may be possible to demonstrate a Fermi arc in NbP by ARPES by observing a kink, criterion (\ref{kink}), but only if the Fermi level can be raised $>$ 20 meV. Lastly, we point out that the counting argument (\ref{Chern}) as recently applied to TaAs \cite{TaAsThem, TaAsChen} and NbP \cite{NbPThem} is invalid because certain Weyl points are too close together.

Single crystal TaAs and NbP samples were grown by chemical vapor transport methods. Angle-resolved photoemission spectroscopy (ARPES) measurements were performed using a Scienta R4000 at Beamline 5-4 of the Stanford Synchrotron Radiation Lightsource, SLAC and Beamline 4 of the Advanced Light Source, LBNL in CA, USA. The angular resolution was better than 0.2$^{\circ}$ and the energy resolution better than 20meV. Samples were cleaved $\textit{in situ}$ and measured under vacuum better than $10^{-10}$ Torr at temperatures $<$ 30K. First principles electronic structure calculations of NbP were carried out with the OpenMX code \cite{OpenMX}, based on the generalized gradient approximation (GGA) \cite{Perdew}. Spin-orbit coupling was incorporated through $j$-dependent pseudo-potentials. For each Nb atom three, three, three and one optimized radial functions were allocated for the s, p, d, and f orbitals ($s3p3d3f1$), respectively. For each P atom, $s3p3d2f1$ were adopted. A $k$-point mesh of $17 \times 17 \times 5$ for the conventional unit cell was used and experimental lattice parameters were adopted in the calculations \cite{Crystal1, Crystal2, Crystal3}. We use Nb $s$ and $d$ orbitals and P $p$ orbitals to construct Wannier functions without performing the procedure for maximizing localization \cite{Wannier1, Wannier2}.  We calculated the surface spectral weight of a semi-infinite (001) slab using an iterative Green's function method based on the Wannier function basis set.

We show that the surface state band structure of NbP as measured by ARPES does not satisfy any of the Fermi arc criteria. We note that vacuum ultraviolet ARPES is sensitive to the surface states of NbP, see Supplementary Information (SI). The surface states consist of lollipop and peanut-shaped pockets and we find that both are hole-like, see Figs. \ref{Fig2}(a)-(d). \textit{Criterion (1)}. Both the lollipop and peanut pockets are closed, so we observe no single disconnected arc. \textit{Criterion (2)}. We see no evidence of a kink in the constant-energy contours, so we do not observe a pair of arcs connecting to the same $W_2$ in a discontinuous way. \textit{Criterion (3)}. We observe an even number of contours everywhere, so no arc. \textit{Criterion (4)}. We can study the Fermi velocity using a difference map of two ARPES spectra, at $E_B = 0.05$ eV and $E_B = 0.1$ eV, shown in Fig. \ref{Fig2}(f). We see that the Fermi velocities have the same sign all the way around both the lollipop and the peanut pockets. If the lollipop consisted of Fermi arcs, one arc should evolve in a hole-like way, while the other arc should evolve in an electron-like way, so the different regions of the lollipop pocket should have opposite Fermi velocities in this sense. Because all points on both pockets have the same Fermi velocity, within the resolution of our ARPES measurements these pockets are trivial, hole-like surface states, and we observe no topological Fermi arcs. We can also consider a closed path through the surface Brillouin zone and count chiralities of edge modes along the path, illustrated in Figs. \ref{Fig2}(e,g). We check a triangular path, $\mathcal{C}$, along $\bar{\Gamma}-\bar{X}-\bar{M}-\bar{\Gamma}$, which encloses net chiral charge $+1$, Fig. \ref{Fig2}(h), and a small circular path, $\mathcal{P}$, which encloses net chiral charge $-2$, Fig. \ref{Fig2}(i). For each path, we label each spinless crossing with an up or down arrow to indicate the sign of the Fermi velocity. We find that going around either $\mathcal{C}$ and $\mathcal{P}$ we have net zero chirality, showing zero Chern number on the associated bulk manifold. The surface states of NbP fail all criteria for Fermi arcs. We cannot show that NbP is a Weyl semimetal using only the surface state band structure from ARPES.

Next, we compare our experimental results to numerical calculations on NbP and show that it is challenging to observe Fermi arcs in our spectra because of the low spin-orbit coupling. We present a calculation of the (001) surface states in NbP for the P termination, at the binding energy of $W_2$, $\varepsilon_{W2} = -0.026$ eV, at the Fermi level and at the binding energy of $W_1$, $\varepsilon_{W1} = 0.053$ eV, see Figs. \ref{Fig3}(a)-(c). We also plot the Weyl point projections, obtained from a bulk band structure calculation \cite{FourCompounds}. We observe surface states (1) near the mid-point of the $\bar{\Gamma}-\bar{X}$ and $\bar{\Gamma}-\bar{Y}$ lines and (2) near $\bar{Y}$ and $\bar{X}$. The surface states (1) form two Fermi arcs and two closed contours at $\varepsilon_{W2}$, see Fig. \ref{Fig3}(d),(e). These states undergo a Lifshitz transition near $\varepsilon_F$ with the surface states (2), forming a large hole-like pocket below the Fermi level. The surface states (2) also form a large hole-like pocket. They contain within them, near the $\bar{X}$ and $\bar{Y}$ points, a short Fermi arc connecting each pair of $W_1$, see Fig. \ref{Fig3}(f),(g). We note the excellent agreement with our ARPES spectra, where we also see lollipop and peanut contours which evolve into trivial, closed, hole-like pockets below $\varepsilon_F$. We also find in our calculation that the separation of Weyl points and the surface state spin-splitting is small. This result is consistent with our ARPES spectra, which do not show spin-splitting in the surface states near the Fermi level.

The small spin-splitting observed in our numerical calculations underlines the difficulty in observing topological Fermi arc surface states in NbP. The separation of the Weyl points is $< 0.01 \textrm{\AA}^{-1}$ for the $W_1$ and $< 0.02 \textrm{\AA}^{-1}$ for the $W_2$, both well below the typical linewidth of our ARPES spectra, $\sim 0.05 \textrm{\AA}^{-1}$. For this reason, we cannot resolve the momentum space region between the $W_1$ or the $W_2$ to determine if there is an arc. We emphasize that we cannot surmount this difficulty by considering Fermi level crossings on $\mathcal{P}$ or $\mathcal{C}$, as shown in Fig. \ref{Fig2}(e). It is obvious that if we cannot resolve the two Weyl points in a Fermi surface mapping, then we also cannot resolve a Fermi arc connecting them in a cut passing through the Weyl points. In this way, on $\mathcal{P}$ we cannot verify the arc connecting the $W_1$ and on $\mathcal{P}$ and $\mathcal{C}$ we cannot verify the empty region between the $W_2$. As an additional complication, it is difficult to use $\mathcal{P}$ or $\mathcal{C}$ because the Fermi level is below the Lifshitz transition for the $W_2$ in NbP, see Fig. \ref{Fig4}a. This invalidates use of criterion (4) because for $\varepsilon_F < \varepsilon_L$, there is no accessible binding energy where the bulk band structure is gapped along an entire loop passing in between a pair of $W_2$, illustrated in Fig. \ref{Fig4}b by the broken dotted red line for $\varepsilon = \varepsilon_F$. However, if we could access $\varepsilon > \varepsilon_L$, then it may be possible to use criterion (2) to demonstrate a Fermi arc in NbP without resolving the $W_2$ and counting chiralities. In particular, while the Fermi arc may appear to form a closed contour due to the small separation of $W_2$, it could have a kink at the location of the $W_2$. Unlike criterion (4), applying criterion (2) to (001) NbP would not depend on resolving the $W_2$ or the spin-splitting of the surface states. Improvements in the quality of NbP crystals or the cleaved surface could also allow the Fermi arcs to be resolved by reducing the spectral linewidth. Lastly, we point out that our results invalidate recent claims that $\mathcal{C}$ \cite{NbPThem} or any direct measurement of Fermi arcs \cite{NbPAndo, NbPChen} can be used to demonstrate a Weyl semimetal in NbP, due to the large linewidth of available ARPES spectra.

We apply criterion (4) to demonstrate Fermi arcs in TaAs. The larger spin-splitting in TaAs increases the separation of Weyl points as compared to NbP, see Fig. \ref{Fig4}c. However, we emphasize that only the $W_2$ are well within experimental resolution. The separation of the $W_1$ in TaAs is comparable to the separation of the $W_2$ in NbP, which as noted above, cannot be resolved. In contrast to NbP, the (001) surface states of TaAs near $\bar{\Gamma}$ consist of crescent pockets with clear spin-splitting, see Fig. \ref{Fig4}(d). We also observe bowtie surface states near $\bar{X}$ and $\bar{Y}$. We apply criterion (4) to search for Fermi arcs in TaAs on paths $\mathcal{C}$ and $\mathcal{P}$, shown in Fig. \ref{Fig4}(e). We mark the Fermi velocity of each crossing with an arrow, but now with one arrow per spinful crossing. Along $\mathcal{C}$, see Fig. \ref{Fig4}(f), we see two well-resolved crossings not far from $\bar{\Gamma}$. However, the spinful crossings from the bowtie pocket are difficult to resolve. Based on the constant-energy contour, Fig. \ref{Fig4}(d), we may interpret the bowtie pocket as consisting of two slightly-separated spinful contours, so we associate two arrows with each of the remaining two crossings along $\mathcal{C}$. We find that the enclosed Chern number is $+2$, while it is $+3$ according to a numerical calculation of the band structure \cite{FourCompounds}. Again, this inconsistency results from the small separation of the $W_1$. As a result, we cannot resolve the additional crossing from the Fermi arc connecting the $W_1$ near $\bar{X}$, which would give a Chern number of $+3$. We can avoid the bowtie pocket by using $\mathcal{P}$, see Fig. \ref{Fig4}g. Here, we see only the two well-separated states near $\bar{\Gamma}$. We find two edge modes of the same chirality, unambiguously showing a Chern number $+2$ on the associated bulk manifold, satisfying criterion (4) for a Weyl semimetal. In this way, we demonstrate that TaAs is a Weyl semimetal by studying only the surface states as measured by ARPES. We note that our results invalidate earlier measurements on $\mathcal{C}$ used to demonstrate a Weyl semimetal in TaAs \cite{TaAsThem, TaAsChen}. Here, we have shown that the small separation between the $W_1$ in TaAs makes it impossible to calculate the Chern number on $\mathcal{C}$. In the same way, our results on NbP invalidate recent works claiming a Weyl semimetal in NbP using $\mathcal{C}$ \cite{NbPThem} or by directly observing Fermi arcs \cite{NbPAndo, NbPChen}. We emphasize that to show a Weyl semimetal it is not enough to present an overall agreement between ARPES and numerics. In the case of TaAs and NbP, there are many trivial surface states, so an overall agreement is not entirely relevant for the topological invariants or the Weyl semimetal state. In addition, an overall agreement is precarious in cases where the system is near a critical point or where there may be several closely-related crystal structures. Rather, to show a Weyl semimetal, it is sufficient to pinpoint a topological Fermi arc in an ARPES spectrum of surface states. Here, we have presented a set of general and distinct criteria, applicable to any material, any one of which pinpoints a topological Fermi arc. By presenting criteria for Fermi arcs, we provide a useful reference for demonstrating novel Weyl semimetals.

\vspace{0.2cm}

{\bf Acknowledgments.} Work at Princeton University and Princeton-led synchrotron-based ARPES measurements are supported by the Emergent Phenomena in Quantum Systems Initiative of the Gordon and Betty Moore Foundation under grant GBMF4547 (M.Z.H.). Single-crystal growth is supported by the National Basic Research Program of China under grants 2013CB921901 and 2014CB239302 and characterization is supported by the U.S. Department of Energy, Office of Science, Basic Energy Sciences under grant DE-FG-02-05ER46200. H.L. acknowledges the Singapore National Research Foundation under award NRF-NRFF2013-03. The work at Northeastern University is supported by the U.S. Department of Energy, Office of Science, Basic Energy Sciences under grant DE-FG02-07ER46352, and benefited from theory support at the Advanced Light Source and the allocation of supercomputer time at the National Energy Research Scientific Computing Center under U.S. DOE grant DE-AC02-05CH11231. I.B. acknowledges the support of the U.S. National Science Foundation GRFP. M.N. is supported by the Los Alamos National Laboratory LDRD program and start-up funds from the University of Central Florida. We thank Makoto Hashimoto and Donghui Lu for technical assistance with ARPES measurements at SSRL Beamline 5-4, SLAC, Menlo Park, CA, USA. We also thank Nicholas Plumb and Ming Shi for technical assistance with ARPES measurements at the HRPES endstation of the SIS beamline, Swiss Light Source, Villigen, Switzerland. I.B. thanks Titus Neupert for useful discussions.


\begin{thebibliography}{99}

\bibitem{Weyl} H. Weyl. Z. Phys. {\bf 56}, 330 (1929).
\bibitem{Herring} C. Herring. Phys. Rev. {\bf 52}, 365 (1937).
\bibitem{Abrikosov} A. A. Abrikosov \& S. D. Beneslavskii. J. Low Temperature Physics {\bf 5}, 141 (1971).
\bibitem{Nielsen} H. B. Nielsen \& M. Ninomiya. Phys. Lett. B {\bf 130}, 389 (1983).
\bibitem{Volovik} G. E. Volovik, \textit{The Universe in a Helium Droplet} (Clarendon Press, Oxford, 2003).
\bibitem{Murakami} S. Murakami. New Journal of Physics {\bf 9}, 356 (2007).
\bibitem{Multilayer} A. A. Burkov \& L. Balents. Phys. Rev. Lett. {\bf 107}, 127205 (2011).
\bibitem{Pyrochlore} X. Wan, A. M. Turner, A. Vishwanath \& S. Y. Savrasov. Phys. Rev. B {\bf 83}, 205101 (2011).
\bibitem{Vish} A. Turner \& A. Vishwanath. arXiv:1301.0330.
\bibitem{Hosur} P. Hosur \& X. Qi. Comp. Rend. Phy. {\bf 14}, 857 (2013).
\bibitem{TaAsThyUs} S.-M. Huang, S.-Y. Xu, I. Belopolski, C.-C. Lee, G. Chang, B. K. Wang, N. Alidoust, G. Bian, M. Neupane, C. Zhang, S. Jia, A. Bansil, H. Lin \& M. Z. Hasan. Nat. Commun. {\bf 6}, 7373 (2015).
\bibitem{TaAsThyThem} H. Weng, C. Fang, Z. Fang, B. A. Bernevig \& X. Dai. Phys. Rev. X {\bf 5}, 011029 (2015).
\bibitem{TaAsUs} S.-Y. Xu, I. Belopolski, N. Alidoust, M. Neupane, G. Bian, C. Zhang, R. Sankar, G. Chang, Z. Yuan, C.-C. Lee, S.-M. Huang, H. Zheng, J. Ma, D. S. Sanchez, B. K. Wang, A. Bansil, F. C. Chou, P. P. Shibayev, H. Lin, S. Jia \& M. Z. Hasan. Science {\bf 349}, 613 (2015).
\bibitem{LingLu} L. Lu, Z. Wang, D. Ye, L. Ran, L. Fu, J. D. Joannopoulos \& M. Solja\v{c}i\'{c}. Science {\bf 349}, 622 (2015).
\bibitem{TaAsThem} B. Q. Lv, H. M. Weng, B. B. Fu, X. P. Wang, H. Miao, J. Ma, P. Richard, X. C. Huang, L. X. Zhao, G. F. Chen, Z. Fang, X. Dai, T. Qian \& H. Ding. Phys. Rev. X {\bf 5}, 031013 (2015).
\bibitem{TaAsNodesDing} B. Q. Lv, N. Xu, H. M. Weng, J. Z. Ma, P. Richard, X. C. Huang, L. X. Zhao, G. F. Chen, C. E. Matt, F. Bisti, V. N. Strocov, J. Mesot, Z. Fang, X. Dai, T. Qian, M. Shi \& H. Ding. Nat. Phys. {\bf 11}, 724 (2015).
\bibitem{NbAs} S.-Y. Xu, N. Alidoust, I. Belopolski, Z. Yuan, G. Bian, T.-R. Chang, H. Zheng, V. N. Strocov, D. S. Sanchez, G. Chang, C. Zhang, D. Mou, Y. Wu, L. Huang, C.-C. Lee, S.-M. Huang, B. K. Wang, A. Bansil, H.-T. Jeng, T. Neupert, A. Kaminski, H. Lin, S. Jia \& M. Z. Hasan. Nat. Phys. {\bf 11}, 748 (2015).
\bibitem{TaPUs} S.-Y. Xu, I. Belopolski, D. S. Sanchez, C. Zhang, G. Chang, C. Guo, G. Bian, Z. Yuan, H. Lu, T.-R. Chang, P. P. Shibayev, M. L. Prokopovych, N. Alidoust, H. Zheng, C.-C. Lee, S.-M. Huang, R. Sankar, F. C. Chou, C.-H. Hsu, H.-T. Jeng, A. Bansil, T. Neupert, V. N. Strocov, H. Lin, S. Jia \& M. Z. Hasan. Sci. Adv. {\bf 1}, 10 (2015).
\bibitem{TaPThem} N. Xu, H. M. Weng, B. Q. Lv, C. Matt, J. Park, F. Bisti, V. N. Strocov, D. Gawryluk, E. Pomjakushina, K. Conder, N. C. Plumb, M. Radovic, G. Aut\`{e}s, O. V. Yazyev, Z. Fang, X. Dai, G. Aeppli, T. Qian, J. Mesot, H. Ding \& M. Shi. arXiv:1507.03983.
\bibitem{TaAsChen} L. X. Yang, Z. K. Liu, Y. Sun, H. Peng, H. F. Yang, T. Zhang, B. Zhou, Y. Zhang, Y. F. Guo, M. Rahn, D. Prabhakaran, Z. Hussain, S.-K. Mo, C. Felser, B. Yan \& Y. L. Chen. Nat. Phys. {\bf 11}, 728 (2015).
\bibitem{HaoNbP} H. Zheng, S.-Y. Xu, G. Bian, C. Guo, G. Chang, D. S. Sanchez, I. Belopolski, C.-C. Lee, S.-M. Huang, X. Zhang, R. Sankar, N. Alidoust, T.-R. Chang, F. Wu, T. Neupert, F. C. Chou, H.-T. Jeng, N. Yao, A. Bansil, S. Jia, H. Lin \& M. Z. Hasan. ACS Nano. DOI:10.1021/acsnano.5b06807 (2015).
\bibitem{Bernevig} A. Bernevig \& T. Hughes, \textit{Topological Insulators and Topological Superconductors} (Princeton University Press, Princeton, 2013).
\bibitem{Na3Bi} S.-Y. Xu, C. Liu, S. K. Kushwaha, R. Sankar, J. W. Krizan, I. Belopolski, M. Neupane, G. Bian, N. Alidoust, T.-R. Chang, H.-T. Jeng, C.-Y. Huang, W.-F. Tsai, H. Lin, P. P. Shibayev, F. C. Chou, R. J. Cava \& M. Z. Hasan. Science {\bf 347}, 294 (2015).
\bibitem{NbPThem} D.-F. Xu, Y.-P. Du, Z. Wang, Y.-P. Li, X.-H. Niu, Q. Yao, P. Dudin, Z.-A. Xu, X.-G. Wan \& D.-L. Feng. Chin. Phys. Lett. {\bf 32}, 10 (2015).
\bibitem{NbPAndo} S. Souma, Z. Wang, H. Kotaka, T. Sato, K. Nakayama, Y. Tanaka, H. Kimizuka, T. Takahashi, K. Yamauchi, T. Oguchi, K. Segawa \& Y. Ando. arXiv:1510.01503.
\bibitem{NbPChen} Z. K. Liu, L. X. Yang, Y. Sun, T. Zhang, H. Peng, H. F. Yang, C. Chen, Y. Zhang, Y. F. Guo, D. Prabhakaran, M. Schmidt, Z. Hussain, S.-K. Mo, C. Felser, B. Yan \& Y. L. Chen. Nat. Mat. (2015).
\bibitem{OpenMX} T. Ozaki. Phys. Rev. B {\bf 67}, 155108 (2003).
\bibitem{Perdew} J. P. Perdew, K. Burke \& M. Ernzerhof. Phys. Rev. Lett. {\bf 77}, 3865 (1996).
\bibitem{Crystal1} S. Rundqvist. Nature {\bf 211}, 847 (1966).
\bibitem{Crystal2} J. O. Willerstr\"{o}m. J. Less Common Metals {\bf 99}, 273 (1984).
\bibitem{Crystal3} J. Xu, M. Greenblatt, T. Emge \& P. H\"{o}hn. Inorgan. Chem. {\bf 35}, 845 (1996).
\bibitem{Wannier1} N. Marzari \& D. Vanderbilt. Phys. Rev. B {\bf 56}, 12847 (1997).
\bibitem{Wannier2} A. A. Mostofi, J. R. Yates, Y.-S. Lee, I. Souza, D. Vanderbilt \& N. Marzari. Comp. Phys. Commun. {\bf 178}, 685 (2008).
\bibitem{FourCompounds} C.-C. Lee, S.-Y. Xu, S.-M. Huang, D. S. Sanchez, I. Belopolski, G. Chang, G. Bian, N. Alidoust, H. Zheng, M. Neupane, B. K. Wang, A. Bansil, M. Z. Hasan \& H. Lin. Phys. Rev. B {\bf 92}, 235104 (2015).
\end{thebibliography}
\end{document}


\title{Supplementary Information: Criteria for directly detecting topological Fermi arcs in Weyl semimetals}

\author{Ilya Belopolski$^*$} 
\author{Su-Yang Xu$^*$} 
\author{Daniel S. Sanchez$^*$} 
\author{Guoqing Chang} 
\author{Cheng Guo} 
\author{Madhab Neupane} 
\author{Hao Zheng} 
\author{Chi-Cheng Lee} 
\author{Shin-Ming Huang} 
\author{Guang Bian} 
\author{Nasser Alidoust} 
\author{Tay-Rong Chang} 
\author{BaoKai Wang} 
\author{Xiao Zhang} 
\author{Arun Bansil}
\author{Horng-Tay Jeng} 
\author{Hsin Lin} 
\author{Shuang Jia} 
\author{M. Zahid Hasan} 

\pacs{}

\date{\today}

\maketitle

In this supplementary information section, we first present an overview of the crystal structure and electronic band structure of NbP. Then, we argue that vacuum ultraviolet ARPES is sensitive to the surface states of (001) NbP. Lastly, we discuss spin-splitting and phenomena arising from the different orbital contributions of the surface states.

\bigskip
{\bf $\S 1.$ Overview of the crystal, electronic structure of the Weyl semimetal NbP}
\bigskip

Niobium phosphide (NbP) crystallizes in a body-centered tetragonal Bravais lattice, in point group $C_{4v}$ ($4mm$), space group $I4_1md$ (109), isostructural to TaAs, TaP and NbAs \cite{Crystal1, Crystal2, Crystal3}. The crystal structure can be understood as a stack of alternating Nb and P square lattice layers, see Fig. \ref{Fig1}a. Each layer is shifted with respect to the one below it by half an in-plane lattice constant, $a/2$, in either the $\hat{x}$ or $\hat{y}$ direction. The crystal structure can also be understood as arising from intertwined helices of Nb and P atoms which are copied in-plane to form square lattices, with one Nb (or P) atom at every $\pi/2$ rad along the helix. The conventional unit cell consisting of one period of the helices is shown in Fig. \ref{Fig1}b. This helical structure is related to the non-symmorphic $C_4$ symmetry, where a $C_4$ rotation followed by a translation by $c/4$ is required to take the crystal back into itself. We note that NbP has no inversion symmetry, so that all bands are generically singly-degenerate. This is a crucial requirement for NbP to be a Weyl semimetal. We show a photograph of the sample taken through an optical microscope, suggesting that it is a single crystal and of high quality, in Fig. \ref{Fig1}c. A scanning tunneling microscopy (STM) topography of the sample shows a square lattice surface, demonstrating that NbP cleaves on the (001) plane, see Fig. \ref{Fig1}d. The lack of defects further suggests the high quality of the single crystals. From the ionic model, we expect that the conduction and valence bands in NbP arise from Nb $4d$ and P $3p$ orbitals, respectively. However, an \textit{ab initio} bulk band structure calculation along high-symmetry lines shows that NbP does not have a full gap but is instead a semimetal, see $\Sigma-\Gamma$, $Z-\Sigma'$, $\Sigma'-N$ in Fig. \ref{Fig1}e, with the bulk Brillouin zone in Fig. \ref{Fig1}f. In the absence of spin-orbit coupling, the band structure near the Fermi level consists of four Dirac lines, shown in purple in Fig. \ref{Fig1}g, see also \cite{FourCompounds}. These Dirac lines are protected by two vertical mirror planes, shown in blue. After spin-orbit coupling is included, each Dirac line vaporizes into six Weyl points shifted slightly off the mirror plane, marked by the dots in Fig. \ref{Fig1}g. Two Weyl points are on the $k_z = 0$ plane, shown in red, and we call these Weyl points $W_1$. The other four, we call $W_2$. We note that on the (001) surface, two $W_2$ of the same chirality project onto the same point of the surface Brillouin zone, giving rise to a projected Weyl point of chiral charge $\pm 2$. The $W_1$ give projections of chiral charge $\pm 1$, see Fig. \ref{Fig1}h and again \cite{FourCompounds}. We also point out that the chiralities of the $W_1$ are flipped between NbP and TaAs, compare main text Figs. 2g and 4e. This is because the Dirac line is larger in NbP and crosses over the edge of the first Brillouin zone, into the second Brillouin zone. As a result, some of the Weyl points arising from the Dirac line end up in the second Brillouin zone, while Weyl points from the Dirac line of the second Brillouin zone show up in the first Brillouin zone. This flips the chiralities of the $W_1$ between NbP and TaAs. This can also be seen from main text Fig. 4c. See also Fig. 1H of Ref. \cite{TaPUs}.

\bigskip
{\bf $\S 2.$ The surface states of (001) NbP by vacuum ultraviolet ARPES}
\bigskip

Here, we show that we observe surface states but not bulk states in vacuum ultraviolet ARPES on the (001) surface of NbP. In our ARPES spectra, we observe a Fermi surface consisting of lollipop-shaped pockets along the $\bar{\Gamma} - \bar{X}$ and $\bar{\Gamma} - \bar{Y}$ lines and peanut-shaped pockets on the $\bar{M} - \bar{X}$ and $\bar{M} - \bar{Y}$ lines, see Fig. 2(a)-(d) in main text. The spectra are consistent with spectra of surface states of other compounds in the same family, suggesting that these pockets are surface states rather than bulk states. Because $C_4$ symmetry is implemented as a screw axis in NbP, the (001) surface breaks $C_4$ symmetry and the surface state dispersion is not $C_4$ symmetric. Our data suggest that the peanut pockets at $\bar{X}$ and $\bar{Y}$ differ slightly, showing a $C_4$ breaking that demonstrates surface states. We note, however, that this effect is much weaker than in TaAs or TaP \cite{TaAsUs, TaPUs}. This result could be explained by reduced coupling between the square lattice layers, restoring the $C_4$ symmetry of each individual layer. In particular, we note that the lattice constants of NbP are comparable to those of TaAs, while the atomic orbitals are smaller due to the lower atomic number. We expect this effect to be particularly important for surface states derived from the $p_x$, $p_y$, $d_{xy}$ and $d_{x^2-y^2}$ orbitals and indeed we observe no $C_4$ breaking at all for the lollipop pockets, which arise from the in-plane orbitals \cite{FourCompounds}. We conclude that we observe the surface states of NbP in our ARPES spectra. 

\bigskip
{\bf $\S 3.$ Spin-splitting and orbital contributions of the surface states of NbP}
\bigskip

Here, we point out several other features of the (001) surface states of NbP. First, we note that we can observe a spin splitting in the lollipop pocket at $E_B \sim 0.2$ eV, shown in Fig. \ref{Fig4}a and repeated with guides to the eye in Fig. \ref{Fig4}b. Next, we observe in our ARPES spectra that the lollipop and peanut pockets do not hybridize anywhere in the surface Brillouin zone. We present a set of dispersions near the intersection of the lollipop and peanut pockets and we find no avoided crossings, see Fig. \ref{Fig4}c for the locations of the cuts, shown in Fig. \ref{Fig4}e. As mentioned above, we attribute this effect to the different orbital character of the two pockets. In particular, the lollipop pocket arises mostly from in-plane $p$ and $d$ orbitals and the peanut pocket arises mostly from out-of-plane $p$ and $d$ orbitals. The suppressed hybridization may be related to the $C_2$ symmetry of the (001) surface. In particular, we note that the in-plane and out-of-plane orbitals transform under different representations of $C_2$. The contributions from different, unhybridized orbitals to the surface states in NbP may give rise to novel phenomena. For example, we propose that quasiparticle interference between the lollipop and peanut pockets will be suppressed in an STM experiment on (001) NbP. Also, the rich surface state structure may explain why the bulk band structure of NbP is invisible to vacuum ultraviolet ARPES. Specifically, because the surface states take full advantages of all available orbitals, there are no orbitals left near the surface to participate in the bulk band structure. Other phenomena may also arise from the rich surface state structure in NbP. Finally, in Fig. \ref{Fig4}d we present a calculation of the Fermi arcs near the $W_2$ in TaAs, equivalent to the calculation shown in main text Fig. 3d and in agreement with the Fermi arcs we observe by ARPES in TaAs.

\clearpage
\begin{figure*}
\centering
\includegraphics[width=17cm, trim={70 210 70 90}, clip]{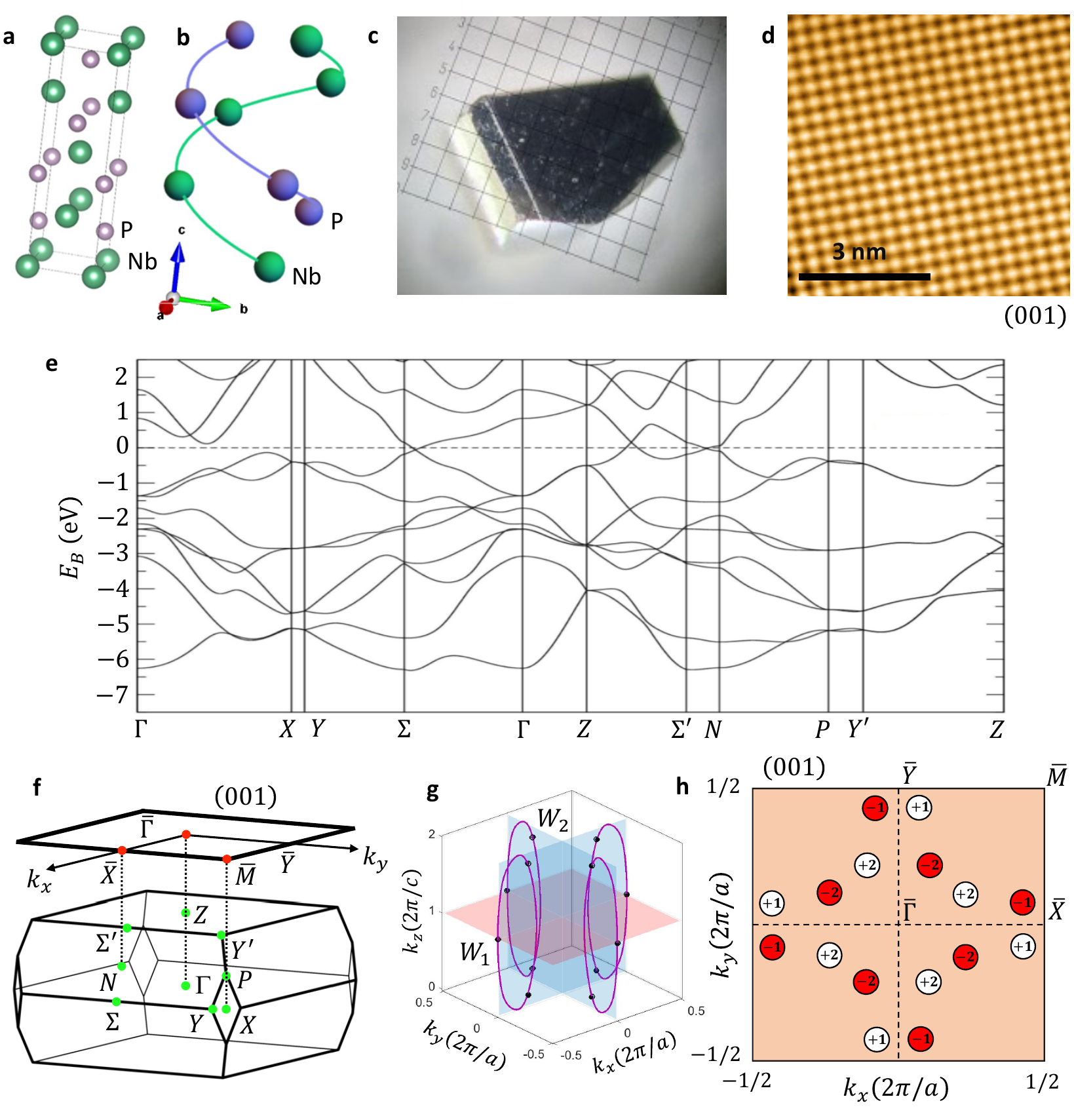}
\end{figure*}

\clearpage
\begin{figure*}
\caption{\label{Fig1}\textbf{Overview of the Weyl semimetal candidate NbP.} (a) The crystal structure of NbP, which can be understood as a stack of square lattices of Nb and P, with a stacking pattern which involves an in-plane shift of each layer relative to the one below it. (b) The crystal structure can also be understood as a pair of intertwined helices of Nb and P atoms which are copied in-plane to form square lattice layers. The axis of the helix is the $\hat{z}$ direction of the conventional unit cell, the center of the helix is $1/4$ of the way along the diagonal of one plaquette of a square lattice layer, and the radius of the helix is $a/2\sqrt{2}$. (c) Photograph of the sample taken through an optical microscope, showing a beautiful crystal. (d) An STM topography of the (001) surface of NbP, showing the high quality of the sample surface, with no defects within a 6.2 nm $\times$ 6.2 nm window. The image was taken at bias voltage $-0.3$ eV and temperature $4.6$ K. (e) \textit{Ab initio} bulk band structure calculation of NbP, using GGA exchange correlation functionals, without spin-orbit coupling (SOC), showing that NbP is a semimetal with band inversions along $\Sigma-\Gamma$, $Z-\Sigma'$ and $\Sigma'-N$. (f) The bulk Brillouin zone and (001) surface Brillouin zone of NbP, with high-symmetry points labeled. (g) Without SOC, NbP has four Dirac lines protected by two mirror planes (blue). With SOC, each Dirac line vaporizes into six Weyl points (indicated by the dots), two on the $k_z = 0$ plane (red, labelled as $k_z = 1$ and equivalent to $k_z = 0$), called $W_1$, and four away from $k_z = 0$, called $W_2$. Note that the Weyl points have very small separation in momentum space, so that each dot corresponds to two Weyl points, one of each chirality. (h) Illustration of the Weyl point projections in the (001) surface Brillouin zone. Two $W_2$ of the same chiral charge project onto the same point in the surface Brillouin zone, giving Weyl point projections of chiral charge $\pm 2$. The separation is not to scale, but the splitting between pairs of $W_2$ is in fact larger than the splitting between pairs of $W_1$, see also main text Fig. 4(c).}
\end{figure*}

\begin{figure}
\centering
\includegraphics[width=17cm, trim={70 430 120 130}, clip]{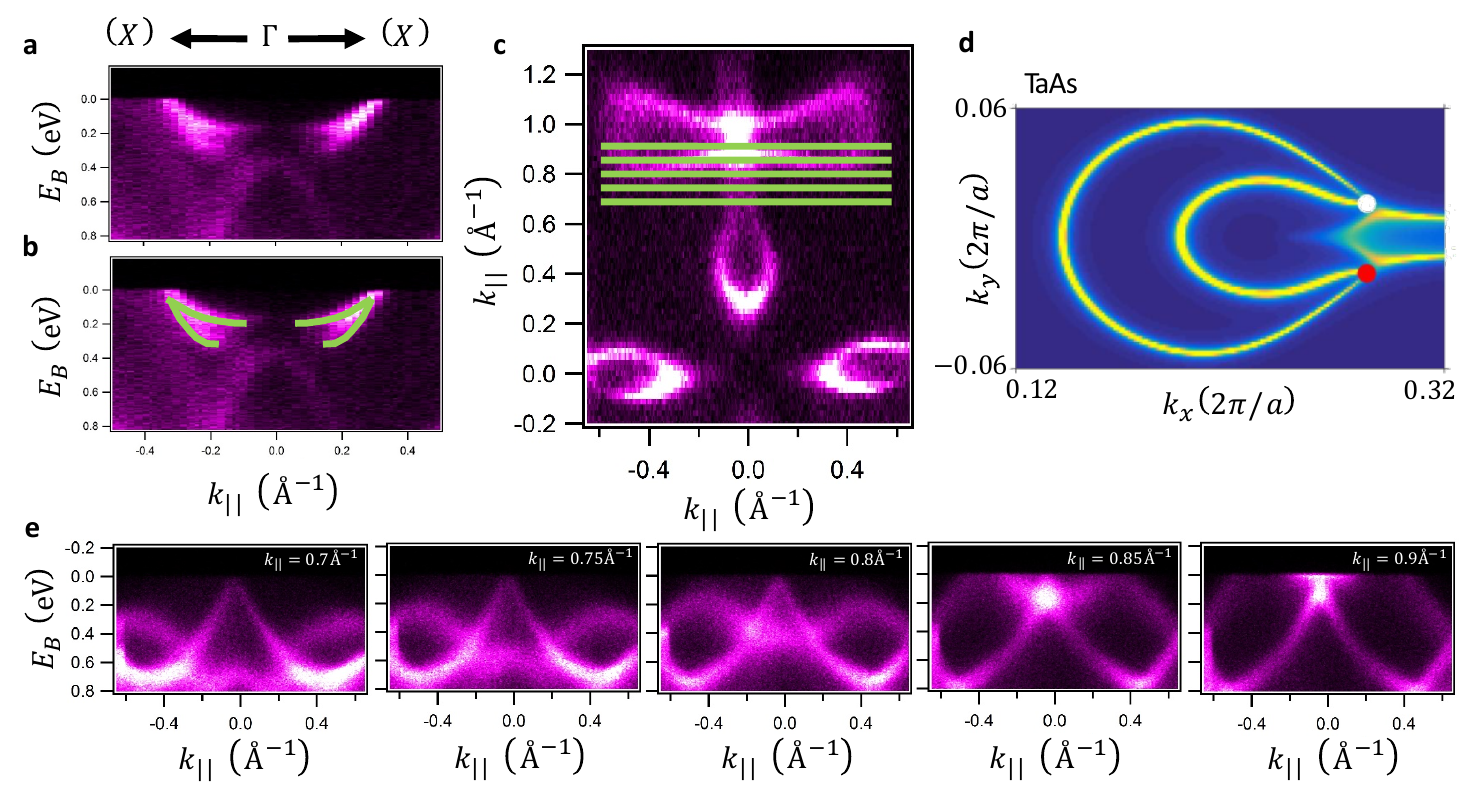}
\caption{\label{Fig4}\textbf{Spin-splitting of the (001) surface states in NbP.} (a) Surface states by ARPES along $\bar{X}-\bar{\Gamma}-\bar{X}$, showing a spin splitting below the Fermi level. (b) Same as (a) but with guides to the eye to mark the spin splitting. (c). Fermi surface by APRES at $E_B = 0.1$ eV, marking the cuts shown in (e). (d) Calculation from \textit{ab initio} of the Fermi arcs near $W_2$ in TaAs, equivalent to main text Fig. 3d for NbP but with much larger spin-splitting between the surface states due to the large spin-orbit coupling. (e) Dispersion of the lollipop and peanut pockets, showing that the two pockets move through each other without any observable hybridization. This absence of an avoided crossing may be related to the different orbital character of the two pockets. Specifically, the lollipop pocket arises from in-plane orbitals and the peanut pocket from out-of-plane orbitals.}

\end{figure}
\clearpage